\PassOptionsToPackage{
    colorlinks=true,
    linkcolor=blue,
    urlcolor=blue,
    citecolor=blue
}{hyperref}

\pdfoutput=1

\documentclass[aps,prd,twocolumn,twoside,superscriptaddress,floatfix,nofootinbib]{revtex4-2}

\usepackage{amsmath,amssymb,gensymb}
\usepackage{graphicx}
\usepackage{bm}
\usepackage{dcolumn}
\usepackage{multirow}
\usepackage{wasysym}
\usepackage{orcidlink}
\usepackage{xcolor}

\setcitestyle{author,open={[},close={]}}

\usepackage[mathlines]{lineno}

\usepackage{hyperref}

\begin{document}

\begin{widetext}
\vspace*{-1.5em}
\noindent\small
© 2026 Optica Publishing Group. One print or electronic copy may be made for personal use only. Systematic reproduction and distribution, duplication of any material in this paper for a fee or for commercial purposes, or modifications of the content of this paper are prohibited.
\vspace*{1em}
\end{widetext}

\title{The Simons Observatory: On-sky performance of radio-transparent multi-layer insulation (RT-MLI) using Styroace-II Styrofoam}

\author{Samuel Day-Weiss\orcidlink{0009-0003-5814-2087}}
\email{dayweiss@princeton.edu}
\affiliation{Joseph Henry Laboratories of Physics, Jadwin Hall, Princeton University, Princeton, NJ, USA 08544}

\author{Nicholas Galitzki\orcidlink{0000-0001-7225-6679}}
\affiliation{Department of Physics, University of Texas at Austin, Austin, TX 78712, USA}
\affiliation{Weinberg Institute for Theoretical Physics, Texas Center for Cosmology and Astroparticle Physics, Austin, TX 78712, USA}

\author{Atsuto Takeuchi}
\affiliation{Department of Physics, The University of Tokyo, Tokyo 113-0033, Japan}

\author{Kam Arnold\orcidlink{0000-0002-3407-5305}}
\affiliation{University of California San Diego, Department of Astronomy and Astrophysics, 9500 Gilman Dr., San Diego, CA  92093-0424}

\author{Kathleen Harrington\orcidlink{0000-0003-1248-9563}}
\affiliation{Argonne National Laboratory, High Energy Physics Division. 9700 S Cass Ave, Lemont, IL 60439}
\affiliation{University of Chicago, Department of Astronomy and Astrophysics. 5801 S Ellis Ave, Chicago, IL 60637}

\author{Masaya Hasegawa\orcidlink{0000-0003-1443-1082}}
\affiliation{High Energy Accelerator Research Organization (KEK), Tsukuba, 305-0801, Japan}

\author{Bradley R. Johnson\orcidlink{0000-0002-6898-8938}}
\affiliation{Department of Astronomy, University of Virginia, Charlottesville, VA 22904, USA}

\author{Akito Kusaka\orcidlink{0009-0004-9631-2451}}
\affiliation{Physics Division, Lawrence Berkeley National Laboratory, Berkeley, CA 94720, USA}
\affiliation{Department of Physics, The University of Tokyo, Tokyo 113-0033, Japan}
\affiliation{Research Center for the Early Universe, School of Science, The University of Tokyo, Tokyo 113-0033, Japan}
\affiliation{Kavli Institute for the Physics and Mathematics of the Universe (WPI), UTIAS, The University of Tokyo, Chiba 277-8583, Japan }

\author{Aashrita Mangu\orcidlink{0009-0000-1028-3524}}
\affiliation{Department of Physics, University of Chicago, 5720 South Ellis Avenue, Chicago, IL 60637, USA}

\author{Jack Orlowski-Scherer\orcidlink{0000-0003-1842-8104}}
\affiliation{Department of Physics and Astronomy, University of Pennsylvania, Philadelphia, PA, 19104, USA}

\author{Lyman A. Page\orcidlink{0000-0002-9828-3525}}
\affiliation{Joseph Henry Laboratories of Physics, Jadwin Hall, Princeton University, Princeton, NJ, USA 08544}

\author{Yoshinori Sueno\orcidlink{0000-0002-3644-2009}}
\affiliation{Joseph Henry Laboratories of Physics, Jadwin Hall, Princeton University, Princeton, NJ, USA 08544}

\author{Osamu Tajima\orcidlink{0000-0003-2439-2611}}
\affiliation{Department of Physics, Faculty of Science, Kyoto University, Kyoto 606-8502, Japan}

\author{Alex Thomas\orcidlink{0000-0001-9528-8147}}
\affiliation{University of Chicago, Department of Astronomy and Astrophysics. 5801 S Ellis Ave, Chicago, IL 60637}
\affiliation{Kavli Institute for Cosmological Physics, University of Chicago, 5640 S Ellis Ave, Chicago, IL, 60637, USA}

\author{Yuhan Wang\orcidlink{0000-0002-8710-0914}}
\affiliation{Department of Physics, Cornell University, Ithaca, NY 14853, USA}

\author{Edward J. Wollack\orcidlink{0000-0002-7567-4451}}
\affiliation{NASA Goddard Space Flight Center, 8800 Greenbelt Road, Greenbelt, MD 20771, USA}

\author{Kyohei Yamada\orcidlink{0000-0003-0221-2130}}
\affiliation{Joseph Henry Laboratories of Physics, Jadwin Hall, Princeton University, Princeton, NJ, USA 08544}

\begin{abstract}
We present the on-sky performance of a Radio-Transparent Multi-Layer Insulation filter (RT-MLI) that uses Styroace-II styrofoam to reject ambient thermal radiation from entering a 0.42\,m diameter aperture to a sub-100\,mK bolometric detector array cooled by a dilution-refrigerator. We find that greater than 90\% of the expected incident infra-red (IR) radiation is rejected, resulting in $<$12\,W of measured transmitted power. Transmitted power in the detector passbands is consistent with a lower bound of 95\%. We address filter design and placement, thermal loading, and mm-wave transmission.

\end{abstract}
\maketitle

\section{\label{sec:intro}Introduction}
The initial deployment of the Simons Observatory (SO) consists of a 6 m crossed-Dragone Large Aperture Telescope (LAT), and a suite of three 0.42 m Small Aperture Telescopes (SATs) \cite{thesimonsobservatorycollaboration2019simonsobservatoryastro2020decadal}. The LAT and SATs are cosmic microwave background (CMB) bolometric polarimeters located in the Parque Astronómico de Atacama on Cerro Toco in Chile, 5200\,m above sea level. The SATs are divided by frequency band, with two instruments at 90/150\,GHz, and a third at 220/280 GHz \cite{Galitzki_2024}. This paper focuses on the former.

\begin{figure}
    \centering
    \includegraphics[width = 0.35\textwidth]{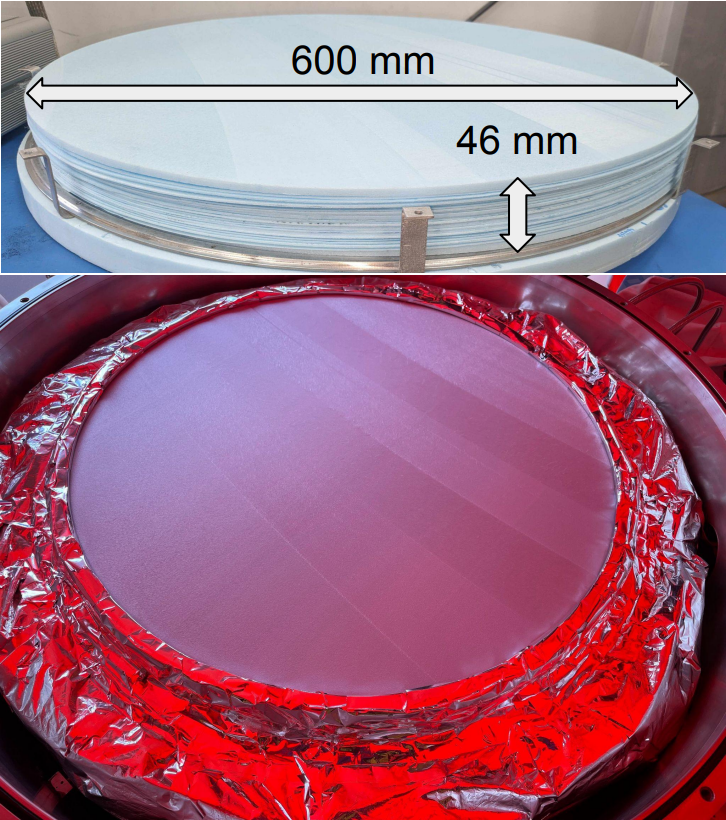}
    \caption{Top: The 24 Styroace-II layers (46 mm thick) shown in the filter holder clamp, flipped upside down onto its sky-facing side. The top side of the small L-brackets shown bolt to the top of the CHWP assembly  (Section \ref{sec:design}). Not shown is the solid aluminum cylinder that aligns the stack above the optics (Figure \ref{fig:rt-mlicad}). The thick piece of foam below the metal holder is not part of the filter stack nor the SAT optics. Bottom: The RT-MLI filter stack installed in the SAT. The flange on the periphery of the photograph is for the ambient vacuum plate that holds the window assembly. The visible MLI blanketing appears red due to reflection of light in the local environment.}
    \label{fig:rt-mliPICs}
\end{figure}

\begin{figure*}
    \centering
    \includegraphics[width = 0.9\textwidth]{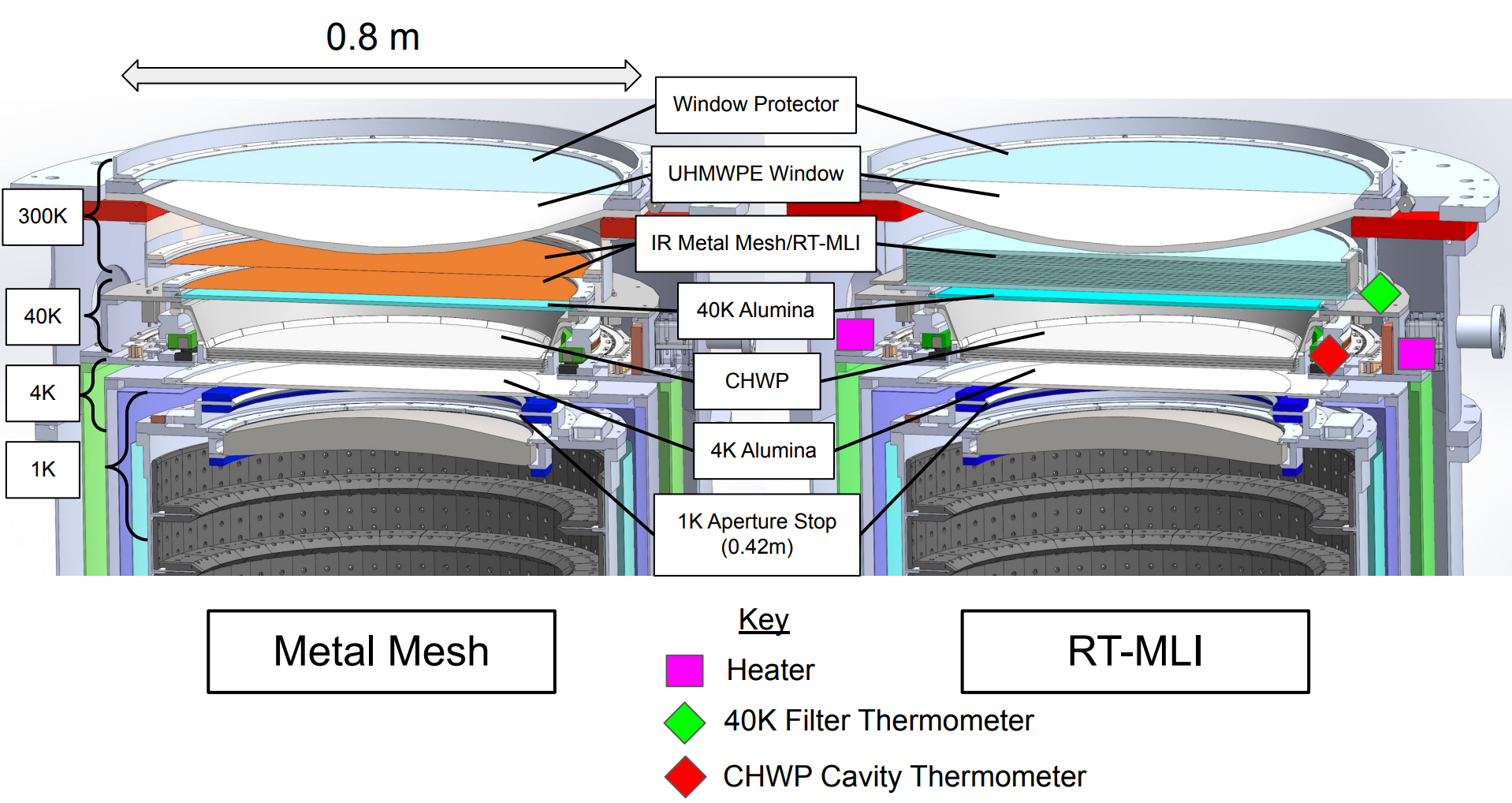}
   \caption{Cross section of the SAT input optics in both the metal mesh and RT-MLI filter configurations. The red plate at the top of the figure forms the primary ambient temperature vacuum seal. The square icons show approximate locations for the 40K alumina filter and CHWP cavity thermometers (see Table \ref{table:irblocking}) used to evaluate the RT-MLI IR rejection performance. The purple rectangles show the heaters that were used for laboratory thermal load curves to simulate radiative loading from the window, and in the on-sky RT-MLI configuration regulate the CHWP assembly temperature to $<$1\,K fluctuations over 12 hours to address possible in-band systematics. In reality there are four heaters distributed evenly around the circumference of the CHWP assembly, but only two are shown. Thermometer and heater locations are consistent across both filter configurations. The MLI used for thermal insulation throughout the cryostat is not shown. Temperatures listed on the left of the figure are approximate.}
    \label{fig:rt-mlicad}
\end{figure*}

The SAT detectors view the millimeter-wave sky via a 0.37\,m$^2$ (690 mm diameter) window, through which we expect  $\sim$120 W of infrared (IR) radiative loading incident on the 40\,K optics of the receiver at the average observing site ambient temperature of 0°C. Primary rejection\textbf{\footnote{In this work, IR rejection refers to the ability of a filter assembly to prevent transmission of ambient radiation to the cryogenic temperature stages. The method of rejection is different for different filter architectures and materials (Section \ref{sec:design}).}} of this radiation is achieved by a Radio-Transparent Multi-Layer Insulation (RT-MLI) filter stack introduced in \citeauthor{Choi:2013mna} \cite{Choi:2013mna} and shown in Figure~\ref{fig:rt-mliPICs}, the performance of which is the focus of this study. While use of RT-MLI-style filters in mm-wave telescopes is well established (e.g., \cite{Ade_2022,sobrin2022design,Lee_2020,inoue2016polarbear}), the details presented here on the in-field performance of a cryogenic (40 K) large diameter Styroace-II styrofoam stack without mechanical or insulating spacers are new. 

 In an idealized RT-MLI filter stack radiative heat transfer is balanced between layers, as well as with the conductive heat transfer between the top and bottom surface of each layer \cite{Choi:2013mna}. In the limit of small layer thickness (i.e., a stack with many thin layers) and assuming that both surfaces above and below the filter stack act as blackbodies, the total power transmitted through the RT-MLI filter can be approximated by the conventional MLI model for cryogenic super-insulation \cite{Choi:2013mna}:
\begin{equation} \label{eq:np1law}
    q_{\mathrm{load}} = \frac{1}{N_{\mathrm{layer}}+1} \cdot \sigma\left[(T^{\mathrm{high}})^4 - (T^{\mathrm{low}})^4\right],
\end{equation}
where the total thermal loading per unit area on the 40 K stage, $q_{\mathrm{load}}$, is calculated from the equilibrium temperatures of the surfaces above and below the filter stack,  $T^{\mathrm{high}}$ (the ambient temperature) and $T^{\mathrm{low}}$, respectively, and $\sigma$, the Stefan-Boltzmann constant. The number of layers in the filter stack is $N_{\mathrm{layer}}$, which in our implementation is 24 ($46\,$mm total thickness). For in-field temperatures, application of Eq. \ref{eq:np1law} yields $\sim$5\,W of power transmitted to the 40 K stage of the SAT.\footnote{This assumes $T^{\mathrm{high}} = 273$K, taken from the average ambient site temperature and $T_{\mathrm{low}} = 44$ K, the measured temperature of the surface below the RT-MLI filter (40 K Alumina filter, Table \ref{table:irblocking}). Variation of the ambient and 40 K bottom surface temperatures at the expected level does not appreciably change the result from Equation \ref{eq:np1law}.} While this model provides order-of-magnitude loading estimates and is a useful guide for filter design, it ignores scattering, and assumes both uniform radial temperature profiles for each layer and that re-radiation only occurs to the sky and surfaces perpendicular to the optical path. We find that Eq. \ref{eq:np1law} underestimates the loading for our configuration, and we propose an additional loading source in Section \ref{IRtherm}.

We measure that the in-field net thermal power transmitted to the cryogenic stages through the RT-MLI filter from the window is $<$12 W. For in-band radiation, \citeauthor{Choi:2013mna} \cite{Choi:2013mna} model a transmittance greater than $0.997^{n}$ for frequencies below 200 GHz, where $n$ is the number of 3\,mm thick layers in the stack. They note that this dependence on $n$ is equivalent to a dependence on total filter thickness. Adapting their model to the filter thickness used in the SAT, we expect a transmission greater than $0.997^{t/3\mathrm{mm}}\sim 0.96$,  where $t = 46\,$mm. Although difficult to evaluate agreement with this model in the deployed SAT configuration, we find that this lower bound is consistent with our measurements of end-to-end efficiency for the full SAT optical stack (Section \ref{inbandperf}).

\section{\label{sec:design} Design}
There are generally three different IR filter stack design philosophies. Significant work has gone into the construction and testing of large-area metal mesh reflective filters with high in-band transmission and fine-tuned reflection characteristics \cite{ade2006review}. However, partially-polarized in-band scattering generated by etching defects in these filters has been found to increase in-band photon loading \cite{Ade_2022}. Absorptive filters reduce scattering, but with large incident power can strain cryocoolers, introduce temperature non-uniformity, and potentially produce significant in-band emission. Lastly, as described above, RT-MLI style filters use a radiative steady-state achieved between the cryostat window and multiple layers of insulating foam. RT-MLI filter materials have high emissivity in the IR. They are also good thermal insulators, and do not conduct heat into their mounting structure, instead reducing the net radiation transfer to lower temperature stages by re-radiating out of the cryostat window. RT-MLI filters achieve radial temperature uniformity via radiative cooling with a common view factor between layers.

Because of their desirable transmission properties, the design and original field configuration of the SAT included a set of metal mesh\footnote{Also called IR shaders.} reflective filters for IR rejection \cite{Galitzki_2024}. Subsequent analysis of instrument performance identified problems with excess instrument polarization (similar to the finding reported in \cite{Ade_2022}), prompting replacement with the RT-MLI style filter. 

Figure~\ref{fig:rt-mlicad} shows the input to the SAT in both the metal mesh \cite{ade2006review} and RT-MLI \cite{Choi:2013mna} IR filter configurations. On the sky-side of the aperture stop, the input optics resemble those deployed in other mm-wave experiments (see, e.g., ACT \cite{swetz2011overview}, ABS \cite{essinger2009atacama}, the BICEP Array \cite{Hui_2018}, and SPT-3G \cite{sobrin2022design}). The first two components form an ambient temperature window assembly: a 15 $\mu$m thick polypropylene film window protector and a 690 mm inner-diameter (ID), 10 mm thick ultra-high molecular weight polyethylene (UHMWPE) window with an expanded teflon anti-reflection (AR) coating.\footnote{Produced by Cardiff University.} Inside the cryostat a cylindrical aluminum baffle protrudes downwards from the window mounting plate to prevent aperture illumination of the volume between the 40 K and 300 K optics. Below the window is the metal mesh \cite{ade2006review}  or RT-MLI style filter stack, followed by a 40 K alumina absorptive filter with two-layer laminate Mullite-Duroid AR coating \cite{Sakaguri_2024}. This IR rejection layer is mounted at the input to the enclosed cryogenic rotating half-wave plate (CHWP) assembly at 40 K, which itself sits above another alumina filter at the 4 K stage and the aperture stop and first lens at 1 K. For the SAT represented in this study, the 4 K absorptive filter as well as the 40 K alumina plates that bookend the CHWP optical stack use a meta-material-based AR coating \cite{Golec_2022}. The other 90/150 GHz instrument uses the Mullite-Duroid coating for these elements. See \cite{Ali_2020,Galitzki_2024,Yamada_2024,Golec_2022} for more details on the SAT cryogenic and optical design.

In the metal mesh configuration (Figure \ref{fig:rt-mlicad}, left) one filter was on the 300 K stage at the end of the cylindrical baffle, followed by a second filter mounted to the CHWP assembly and heat sunk at 40 K. This configuration is similar to that used in ACT \cite{swetz2011overview}.

For the RT-MLI style filter configuration (Figure \ref{fig:rt-mlicad}, right), the SO SATs stack sheets of Styroace-II Styrofoam\footnote{From DuPont, material properties of interest for application in mm-wave telescopes are listed in Table I of \cite{Choi:2013mna}. Other product information is listed \href{https://www.dupontstyro.co.jp/styrofoam/product/styrofoam.php?ac=1&fm=product}{here}. The Styroace-II batch used for this work was manufactured in 2022.} into a single 664 mm ID holder with a 51 mm height, mounted at 40 K on the CHWP assembly. The holder consists of a stock\footnote{\url{https://www.mcmaster.com/4494N39/}} aluminum duct flange mounted on an annular aluminum base plate, with eight evenly spaced adjustable brackets holding down a 4.8 mm thick annular ring to hold the filter stack and accommodate a flexible number of filter layers. The implementation in this study has 24 total layers.\footnote{This updates the design detailed in \cite{Galitzki_2024}. Due to small foam sheet thickness differences in the full filter stack, the number of filter layers differs slightly between the two 90/150 GHz SATs. There is no indication of differing optical transmission or IR rejection between the two instruments.} The top and bottom layers are 6 mm thick to firmly compress the filter stack and prevent potential sagging towards the 40 K alumina filter below. The 22 middle layers are 1.5 mm thick on average. No thermally insulating spacers are used between the layers. Thus, on average, there is roughly 0.2 mm between layers. At ambient temperature the thermal conductivity of Styroace-II Styrofoam (0.028 W/m$\cdot$K \cite{Choi:2013mna}) is more than 500 times smaller than stainless steel, and as such the physical contact between the styrofoam sheets and the aluminum holder, as well as between the sheets themselves, is inconsequential. In all, the RT-MLI filter assembly cost is approximately 1000 USD.

To accommodate the $35^\circ$ field of view, the CHWP and IR filters are placed in a volume with only $\sim$22 cm of vertical clearance, constraining the RT-MLI filter stack to fit within a vertical space of $\sim$5 cm. The IR rejection requirement is set by the need to both remain comfortably within the cooling capacity of the PT-420 pulse tubes, and to maintain the CHWP temperature well below 85 K in order to magnetically levitate the rotation assembly (using YBCO superconductors) \cite{Yamada_2024}. Both are easily achieved (Table \ref{table:irblocking}).

The SAT RT-MLI implementation differs from other currently fielded experiments in combination of material choice, thickness, and temperature stage placement. SPT-3G uses a set of ten 3.17 mm thick HD-30 Zotefoam sheets, separated by G-10 spacers for their receiver observing at 95, 150, and 220 GHz \cite{sobrin2022design}. The BICEP3 receiver, observing at 95 GHz, uses the same thickness and number of Zotefoam sheets as SPT-3G, but glued to and separated by a stack of aluminum rings \cite{Ade_2022} (the BICEP Array \cite{Hui_2018} bases its IR filter design on BICEP3). The filter stacks for both BICEP experiments and SPT-3G are mounted just below the window at ambient temperature. The GroundBIRD experiment, which observes at 145 and 220 GHz, uses Styroace-II as its RT-MLI material, but places it in combination with metal mesh low-pass filters at both the 40 K and 4 K temperature stages \cite{oguri2016groundbird,Lee_2020}. The POLARBEAR-2 cryostats observed at 95 and 150 GHz and also used Styroace-II, mounting the RT-MLI filter at ambient temperature at the back of an 800 mm diameter, 200 mm thick Zotefoam window assembly \cite{inoue2016polarbear}.

\section{\label{sec:perf} Field Performance}
Data for the in-field characterization of the IR filters covers two distinct observing periods.\footnote{Encompassing only a fraction of the SAT's total calendar observing time.} Our evaluation of IR rejection and cryogenic performance compares data from 53 hours of CMB observations in the metal mesh configuration to 277 hours with the RT-MLI filter stack. We evaluate in-band transmission of the RT-MLI filter stack using beam measurements of Jupiter, taken during dedicated planet scans throughout the second observing period. 

\subsection{IR rejection}
 The thermometers that characterize the RT-MLI performance are placed just below the filter stack on the sky-side rim of the 40 K alumina filter, in the CHWP cavity below the 40 K Alumina filter, and on the 40\,K and 4\,K cold heads of both PT-420 pulse tubes  (Figure \ref{fig:rt-mlicad}). The placement of the cryogenic thermometers is consistent between the laboratory and both in-field IR filter configurations.
 
\begin{table} \label{IRtherm}
   \begin{center}   
   \scalebox{0.9}{
   \begin{tabular}{| c|c|c|c |} 
    \hline
    Stage &Lab [K]& Metal mesh [K]& RT-MLI [K]\\
  &@18°C& ($\frac{\mathrm{dP}}{\mathrm{dT}_{\mathrm{amb}}}$ [W/°C])&($\frac{\mathrm{dP}}{\mathrm{dT}_{\mathrm{amb}}}$ [W/°C])\\
 \hline
    40 K cold&30.2& 32.3& 30.9 \\
 head 1 strap& & (1.2)&(0.6)\\ 
  \hline
    40 K cold&29.5& 30.0& 29.1\\
 head 2 (DR)& & (1)&(0.6)\\
 \hline
 40 K Alumina&38.5& N/A&43.7\\
 filter& & &(0.2)\\
 \hline
    CHWP cavity&38.5& 47.8& 44.9\\
 (YBCO holder)& & (0.5)&(0.2)\\
 \hline
\hline
    4 K filter stack &3.1&  4.0 & 3.9\\
 \hline
    4 K cold&2.5& 2.7& 2.7\\
 head 1& & &\\ 
 \hline
    4 K cold&2.4& 2.8& 2.8\\
 head 2 (DR)& & &\\
 \hline
    1 K still &0.832& 0.819& 0.800\\  
    \hline
    0.1 K mixing&0.035& 0.037& 0.040\\
 chamber& & &\\
 \hline
    \end{tabular}
    }
    \end{center}
    \caption{Average temperatures achieved during nominal CMB observations for both filter configurations, scaled to the seasonal observing site average of  0°C. Laboratory temperatures (18°C) achieved with the input 40 K stages shielded from 300 K by aluminum blank-off plates and without a spinning half-wave plate are listed in column two for comparison. In columns 3 and 4 an estimate of the 40 K loading dependence on ambient temperature is provided below the average stage temperature. The 40 K cold head 1 strap entry gives the temperature of a thermometer mounted along the heat strap tying the PT-420 cold head to the cryostat. This sensor reads between $\sim$ 1.2 and 1.7 K greater than the cold head itself, depending on the power through the strap. Temperatures below the double line (below 40K) are not affected by the IR filter. Differences in the 1 K still temperature are accounted for by an unrelated application of different power to the dilution refrigerator (DR) still stage (7.8 mW for metal mesh vs. 5.6 mW for RT-MLI) between observing periods. When the DR is operating within its designed cooling capacity, an increase in still power increases 3He/4He mixture flow rate, and decreases the mixing chamber temperature, which also accounts for the differences at the 0.1\,K stage.}
    \label{table:irblocking}
\end{table}

In the lab, with the 40\,K environment fully shielded (i.e., no IR loading via illumination from the window), the approximate total loading on the 40 K stage based on the PT-420 capacity curve\footnote{\tiny \hyperlink{ https://cdn.bluefors.com/wp-content/uploads/2023/09/22145601/PT420-RM-Capacity-Curve.pdf}{https://cdn.bluefors.com/wp-content/uploads/2023/09/22145601/PT420-RM-Capacity-Curve.pdf}} is 10 W. In the field, the ambient temperature is 18°C colder, reducing this load. The dominant contribution to loading \textit{variation} on the 40 K stage at the observing site is from the change in ambient temperature (diurnal fluctuations).
 
Table \ref{table:irblocking} summarizes the nominal thermal performance in the lab (closed off to the environment) and during CMB observations while using the RT-MLI and reflective metal mesh filter configurations.  Thus, this shows a direct differential comparison between filter configurations. Although the effects are most notable at 40 K, results at the 4\,K,  1\,K, and 0.1\,K stages of the SAT are also reported to give context to the overall thermal performance. Temperatures in columns 3 and 4 are scaled to the seasonal site average ambient temperature of $0^\circ$C. At this temperature we estimate approximately 4 W less loading on the 40 K stage with the RT-MLI filter configuration. This difference is similar to an in-field measurement by BICEP3 of 6 W between their metal mesh and ambient temperature Zotefoam RT-MLI filters \cite{Ade_2022}.

It is clear that the lower 40\,K temperatures achieved in the SAT with the Styroace-II RT-MLI stack make this configuration preferable for IR rejection. The greater susceptibility of the metal mesh stack to ambient temperature fluctuations is also a notable effect, and makes temperature control of the CHWP assembly more challenging (see Figure  \ref{fig:rt-mlicad}). For both filter configurations the propagation of diurnal temperature fluctuations to the colder stages of the SAT is negligible.

To estimate the loading through the window and RT-MLI filter in the field, we reference in-lab load curves taken with a fully-shielded 40K environment, and that used heaters on the 40\,K filter mounting structure to simulate IR loading (Figure \ref{fig:rt-mlicad}). To compare to these curves, we correct for the approximate linear dependence of the 40 K stage temperature on the site ambient temperature to match the laboratory environment at 18°C. Once corrected, the measured in-field temperatures are converted to applied power, thus isolating the loading contribution from opening the window to the 40 K filters. Loading on the 40 K stage from the operation of the CHWP must also be accounted for \cite{Yamada_2024}, which we measure to be $\sim$1\,W. We follow this estimation procedure for all 40 K thermometers present in laboratory testing, which yields an average of 8 $\pm$ 2 W of transmitted IR power. The additional systematic uncertainty in the loading measurement is estimated to be $\sim$15\%, driven by the modeling of the thermal load curves and the measurement of the absolute ambient temperature of the cryostat. We take the maximum measured value of 12 W from the thermometers closest to the filter stack\footnote{The red and green square icons in Figure \ref{fig:rt-mlicad}.} (also the largest value overall) as an upper bound for the transmitted power through the RT-MLI. Based on the measured pulse tube cold head temperatures in Table \ref{table:irblocking} in combination with the standard capacity curve, the 12W is a conservative upper bound, and implies a greater than 90\% rejection of the expected incident IR radiation.

This upper bound is larger than the 5 W predicted by Eq. \ref{eq:np1law}. In our implementation, the cylindrical perimeter of the aluminum RT-MLI holder accounts for 20\% of the total surface area of the filter volume. As described above, the majority of IR loading is re-radiated out of the cryostat window. However, thermal re-radiation transverse to the optical axis that hits the perimeter is partially absorbed and reflected back into the filter stack. This is not accounted for in Eq. \ref{eq:np1law}. With an open, non-reflective filter holder perimeter, \citeauthor{Choi:2013mna} \cite{Choi:2013mna} find the temperature of the bottom layer of a similarly thick styrofoam stack to be 140 K. For our diameter filter, this corresponds to $\sim$4-8 W of transmitted power to the 40 K stage, assuming that the filter acts as a blackbody with emissivity 0.5-1 \cite{Choi:2013mna}. The other 8-4 W would have to be absorbed by the filter holder and environs. If the effect of the reflective perimeter is instead only to increase the temperature of the RT-MLI stack, then the bottom layer would have to be $\sim$155-185 K to radiate the 12 W to the 40 K stage, again assuming emissivities 1-0.5. The reality is likely in between these cases.

\subsection{In-band transmission}\label{inbandperf}

Estimating the in-field transmission of the standalone RT-MLI filter stack is difficult in the context of the full SAT optical assembly. We instead estimate the band-averaged end-to-end efficiency of the optics and detectors using beam measurements of Jupiter (Table \ref{tab:efficiencies}). We assume $T^{\mathrm{Bright.}}_{\mathrm{Jupiter}}= 173$ K for the Jupiter brightness temperature relative to a blank sky. We assume a circular disk shape with a time-varying apparent diameter estimated from the \textit{PyEphem} module.\footnote{\hyperlink{https://rhodesmill.org/pyephem/}{https://rhodesmill.org/pyephem/}} 

The fourth column in Table \ref{tab:efficiencies} summarizes the measured efficiencies for three SAT detector wafers taken during the RT-MLI filter configuration observation window. In total there are seven detector wafers in each SAT, each with 1720 optically-coupled bolometers covering the two polarizations and two frequencies per feed. The reported efficiencies are from the per-observation average of peak fits to Jupiter map beam profiles. For all efficiencies listed in Table \ref{tab:efficiencies} a preliminary in-field measurement of the bandpass of the 90/150 GHz detectors is used. The estimated fractional uncertainty on $\eta_{\mathrm{Jup}}$ is 15\%, which comes from our current knowledge of the passband width and the main beam solid angle. Future work will address both the passband and beam measurements in further detail.

\begin{table}
    \begin{center}
    \begin{tabular}{|c|c|c|c|c|} \hline 
         Wafer & Freq (GHz)& $\#$ obs&  $\eta_{\mathrm{Jup}}$ $[\%]$&  $\eta_{\mathrm{sim}}$ $[\%]$\\ \hline 
         ws0 & 90& 17&  21&  18\\
         &150&21&39&29\\ \hline 
         ws1&90&37&  24&  21\\
         &150&40&31& 28\\ \hline 
         ws6 &90&46&  19&  16\\
         &150&47& 30&27\\ \hline
    \end{tabular}
    \end{center}
    \caption{Jupiter- and simulation-based estimates for the detector wafer averaged end-to-end efficiencies, $\eta$.  The first column is the wafer identifier. Due to Jupiter's availability and position on the sky during the observing period, only three of the SAT's seven wafers are presented. The number of dedicated Jupiter observations for each frequency band is listed in column three. The Jupiter-based measurements (column four) are systematic error dominated, with an estimated fractional uncertainty of 15\%. Column five gives simulated values for the per-wafer end-to-end efficiency assuming a flat transmission of 0.96 for the RT-MLI filter stack. The simulations have an estimated 20\% fractional uncertainty.}
    \label{tab:efficiencies}
\end{table}

Column five of Table \ref{tab:efficiencies} reports a simulated end-to-end efficiency for each detector wafer, $\eta_{\mathrm{sim}}$, estimated from the current model for the SAT detectors and optics, and assuming a flat transmission of 0.96 for the RT-MLI filter. The on-wafer detector efficiencies are given in Dutcher et al. \cite{Dutcher_2023}, modified by an updated estimate for the passband width. For the optical transmission, the simulation uses \textit{jbolo}\footnote{https://github.com/JohnRuhl/jbolo}, an adaptation of the \textit{BoloCalc} pipeline developed by \citeauthor{bolocalc} \cite{bolocalc}. Where available, lab-measured reflection and absorption spectra are used. These are supplemented with expected material properties to cover all components of the SAT optics. The transmission through the atmosphere above the Atacama observing site is also included, modeled with the \textit{am} software \cite{paine_2019_3406483}. A 20\% fractional uncertainty is estimated for $\eta_{\mathrm{sim}}$, primarily due to uncertainty in the SAT passband. End-to-end efficiency estimates from Jupiter are consistent with the model assuming the expected lower bound of 96\% transmission through the RT-MLI stack. An independent, in-lab transmission measurement of the exact SAT Styroace-II filter configuration at room temperature and pressure yields better than 95\% transmission below 200 GHz \cite{WangThomas:private}, consistent with the in-field performance. We adopt the 95\% bound as our baseline.

\section{\label{sec:concl}Conclusion}
We have presented in-field measurements of an RT-MLI IR filter implementation using a compact, low-profile stack of Styroace-II styrofoam mounted at 40 K. The filter configuration meets the cryogenic and optical design criteria for the Simons Observatory Small Aperture Telescopes, and has provided stable IR rejection and in-band transmission throughout four observing seasons (two calendar years) and six cryogenic cycles across the two deployed 90/150 GHz instruments. No deformation of the Styroace-II was observed and IR rejection and in-band transmission performance was recovered through three vacuum pump-out/venting cycles performed in the field at a rate of 5 mBar/minute through $\sim$500 mBar. Moreover, we find that the filter architecture presented is relatively inexpensive to construct, and is straightforward to implement. Further investigation is required to better compare Styroace-II in-band transmission and scattering with that of other foam filters before its use can be recommended at observing bands higher than 150 GHz. We measure that the Styroace-II filter stack rejects more than 90\% of the expected incident out-of-band IR radiation from the cryostat window, while simultaneously demonstrating consistency with a transmittance greater than 95\% of the in-band signal at observing bands centered at 90 and 150 GHz.

\section{Acknowledgements}
This work was supported in part by a grant from the Simons Foundation (Award \#457687, B.K.). This work was supported by the U.S. National Science Foundation (Award Number: \#2153201).Work at LBNL is supported in part by the U.S. Department of Energy, Office of Science, Office of High Energy Physics, under contract No. DE-AC02-05CH11231. This work was supported by Grant-in-Aid for JSPS Fellows. This work was supported in part by World Premier International Research Center Initiative (WPI Initiative), MEXT, Japan. In Japan, this work was supported by JSPS KAKENHI grant Nos. JP19H00674, JP23H00105 and JP24K23938, and the JSPS Core-to-Core Program JPJSCCA20200003. OT acknowledges partial support from JSPS KAKENHI grant number JP22H04913.

\section{Conflicts of Interest}
The authors declare that there are no conflicts of interest related to this article.

\section{Data Availability}
Data analyzed for this study are not publicly available, but may be obtained from the authors upon reasonable request.

\bibliographystyle{elsarticle-num-names}
\bibliography{reference}
\end{document}